\documentclass[iop, apj, twocolappendix, numberedappendix]{emulateapj}

\graphicspath{{./}{figures/}}

\usepackage{graphicx}
\usepackage{dcolumn}
\usepackage{bm}
\usepackage{amsmath}
\usepackage{amssymb}
\usepackage{hyperref}
\usepackage{color}

\newcommand{\refeq}[1]{Equation~(\ref{eq:#1})}

\newcommand{\refEq}[1]{Equation~(\ref{eq:#1})}

\newcommand{\reffig}[1]{Figure~\ref{fig:#1}}

\begin{document}

\title{Standard candles and sirens rescue $H_0$}

%
%
%

\slugcomment{Accepted to \textit{The Astrophysical Journal}: 10/09/2020}

\author{
Aniket Agrawal\altaffilmark{1},
Teppei Okumura\altaffilmark{1,2}, and
Toshifumi Futamase\altaffilmark{3}
}
\affil{
$^1$ Institute of Astronomy and Astrophysics, Academia Sinica, 11F of AS/NTU Astronomy-Mathematics Building, No.1, Sec. 4, Roosevelt Road, Taipei 10617, Taiwan, R.O.C.; aagrawal@asiaa.sinica.edu.tw \\
$^2$ Kavli Institute for the Physics and Mathematics of the Universe (WPI), UTIAS, The University of Tokyo, Kashiwa, Chiba 277-8583, Japan \\
$^3$ Department of Astrophysics and Meteorology, Kyoto Sangyo University, Kita-ku, Kyoto 603-8555, Japan \\
}

\begin{abstract}

We show that future observations of binary neutron star systems with electromagnetic counterparts together with the traditional probes of low- and high-redshift Type Ia supernovae (SNe Ia) can help resolve the Hubble tension. The luminosity distance inferred from these probes and its scatter depend on the underlying cosmology. By using the gravitational lensing of light or gravitational waves emitted by, and peculiar motion of, these systems we derive constraints on the sum of neutrino masses, the equation of state of dark energy parametrized in the form $w_0 + w_a (1-a)$, along with the Hubble constant and cold dark matter density in the universe. We show that even after marginalizing over poorly constrained physical quantities, such as the sum of neutrino masses and the nature of dark energy, low-redshift gravitational-wave observations, in combination with SNe Ia, have the potential to rule out new physics as the underlying cause of the Hubble tension at $\gtrsim 5.5\sigma$. 
\end{abstract}


\section{Introduction} \label{sec:intro}

The standard model of cosmology is under stress. 
There is at least a $4\sigma$ discrepancy between the values of the expansion rate of the universe today, the Hubble constant $h \equiv H_0/ (100$ km s$^{-1}$Mpc$^{-1}$), measured using early and late universe probes~\citep{Verde:2019ivm}. A solution to this discrepancy will either revolutionize our understanding of the physical world or help us discover previously unknown systematics. Therefore, significant theoretical and observational effort is currently underway to uncover the cause of this `Hubble tension'. On the theory side, several models have been proposed to explain the origin of this discrepancy, from early dark energy~\citep{Poulin:2018cxd}, self-interacting neutrinos~\citep{Kreisch:2019yzn}, or scalar fields that inject energy locally around the matter-radiation equality~\citep{Agrawal:2019lmo}, to effects of a local inhomogeneity~\citep{Kasai:2019yqn}. Since there is still disagreement as to the theoretical underpinnings of any of these models, another approach is to make better measurements of $h$ using as many probes as possible. Different systematics affecting different probes allow a check on systematics as an origin of the Hubble tension. It is shown in~\cite{Verde:2019ivm} that low-redshift probes such as Type Ia supernovae (SNe Ia), strong lenses, water masers, and surface brightness fluctuations seem to be consistent with each other and give a value $h \sim 0.73$. High-redshift probes including the cosmic microwave background (CMB) measurements from Planck~\citep{Aghanim:2018eyx} and those obtained from combining galaxy clustering with early universe Big Bang nucleosynthesis also agree with each other and predict $h \sim 0.67$. There is, thus, a disagreement between the values measured at high and low redshift. 

Direct detection of gravitational waves (GWs) has opened up a new window into the universe~\citep{Abbott:2016blz}. By measuring the time variation of GWs we can measure the luminosity distance to their sources. These distances are poised to be measured with extremely high precision. Optical follow-up of sources that also emit light allows us to measure their redshift~\citep{TheLIGOScientific:2017qsa}. Technological limits restrict us to detecting GWs from low-redshift sources only. Nevertheless, they provide a completely independent way to measure $h$ at these redshifts. In this paper we show that by combining future measurements of $h$ using GW sources {out to $z \lesssim 0.1$ and SNe Ia out to $z \lesssim 1.7$}, we can potentially resolve the Hubble tension by providing extremely tight constraints on $h$ in these two redshift ranges. As we show, the two values are expected to disagree by more than $5.5\sigma$ allowing us to conclude that at least some of the Hubble tension originates from systematic uncertainties and not from any physical effects. This conclusion is robust to any assumptions about the nature of dark energy or the sum of neutrino masses that we marginalize over. 

\section{The Magnitude-Redshift relation}\label{sec:mzrelation}

Cosmological information from both standard candles (SNe Ia), and standard sirens (GW sources), is contained in the luminosity distance, $d_{\text{L}}(z) = (1+z)\chi(z)\,,$ where $\chi(z)$ is the comoving distance at the same redshift, 
\begin{align}\label{eq:chi_z}
    \chi(z) = \frac{1}{H_0}\int_0^z dz' \frac{1}{E(z')}\,,
\end{align}
with
\begin{align}\label{eq:ez}
    \nonumber E^2(z)=&\Omega_r (1+z)^4+\Omega_M (1+z)^3+\Omega_K(1+z)^2\\ &+\Omega_{\Lambda}(1+z)^{3(1+w_0+w_a)}e^{-3w_a z/(1+z)}\,,
\end{align}
where $\Omega_r,\, \Omega_M,\, \Omega_K$, and $\Omega_{\Lambda}$ are the energy density fractions of radiation, matter, curvature, and dark energy, respectively, and $w \equiv w_0+w_a(1-a)$ is the time-varying equation of state for dark energy, parameterized by $w_0$, which characterizes the constant part, and $w_a$, which represents the amplitude of time variation~\citep{Chevallier:2000qy,linder:2003}. Conventionally, the observed quantity is the apparent magnitude, $m(z)$, which is related to $d_L(z)$ as $m(z) = 5\, \text{log}_{10} d_{\text{L}}(z)+M \,,$ where $M$ is the absolute magnitude of a source, which can be determined. For SNe Ia $M$ is calibrated using their observed peak luminosity and its subsequent decay, while for GW sources it is calculated using the spectral and temporal variation of the emitted GWs. Thus, by measuring $m(z)$, and thus $d_L(z)$, one can constrain the cosmological parameters. 

These equations for luminosity distance hold for sources that have no peculiar velocity and are observed in a homogeneous universe, such that the emitted electromagnetic or GWs are not gravitationally lensed. In an inhomogeneous universe, matter along the line of sight (l.o.s.) affects the propagation of these waves via lensing and Sachs--Wolfe and integrated Sachs--Wolfe effects~\citep{sachs:1967}. In the matter domination era, lensing is the dominant effect among these. Lensing changes the observed brightness of a given source, making it appear either closer or farther than it actually is, thus changing the observed luminosity distance. 

Peculiar motion of the source affects the observed redshift. As a result, it affects the predicted luminosity distance to the source. The total peculiar motion is a sum of two components---one of cosmological origin, which is sourced by the large-scale structure of the universe, and the other of astrophysical origin, which is sourced by the small-scale dynamics of the host galaxy. Cosmological information can only be gleaned from the first component, which can be measured separately by measuring the peculiar velocity of the galaxy as a whole. For example, if the host galaxy is part of a cluster we need to use the cluster redshift~\citep{Leget:2018juj}. Thus, in the rest of the paper, we consider only cosmological peculiar velocities.  

As shown in~\cite{Hui:2005nm} lensing and peculiar motion respectively alter the inferred luminosity distance by
\begin{align}\label{eq:deltal}
\nonumber\delta d_{\text{L, lens}}(z_s, \hat{\textbf{n}}) =&-\frac{3 H^2_0 \Omega_{m0}}{2}\int_{0}^{\chi_s}d\chi \frac{\chi(\chi_s-\chi)}{\chi_s}\\
&\times (1+z)\delta_m(z, \hat{\textbf{n}})\,, \\
\label{eq:deltav}
\delta d_{\text{L, vel}}(z_s, \hat{\textbf{n}}) =& \Bigg[\frac{\textbf{v}_{o}\cdot\hat{\textbf{n}}-\textbf{v}_{s}\cdot\hat{\textbf{n}}}{a_s H_s\chi_s}\Bigg]+\textbf{v}_{s}\cdot\hat{\textbf{n}}\,.
\end{align}

Here $\hat{\textbf{n}}$ is the unit vector in the \emph{observed} l.o.s. direction, $\chi_s$ is the comoving distance at \emph{observed} redshift $z_s$ of the source, $a_s$ is the scale factor corresponding to $z_s$, $H_s$ is the Hubble rate at redshift $z_s$, $\delta_m(z, \hat{\textbf{n}})$ is the matter density fluctuation at redshift $z(\chi)$ in the direction $\hat{\textbf{n}}$, and $\textbf{v}_{o}$ and $\textbf{v}_{s}$ are the peculiar velocities of the observer and the host galaxy, respectively. These equations follow from linearizing the Einstein equation for small metric perturbations, so that terms of higher order can be dropped. Despite the linear approximation it can still be used to account for some nonlinearity in density and velocity perturbations because they are second and first derivatives of the metric perturbation, respectively (assuming a linear relation between density and velocity). It is also noteworthy that propagating GWs obey the \emph{same} equations as photons, as long as their amplitude is small and their wavelength is long enough so that the ray optics limit is realized~\citep{Misner:1974qy}. 

The observed magnitude changes by~\citep{Hada:2014jra, Hada:2016dje, Hada:2018ybu}
\begin{align}\label{eq:delm_deld}
\delta m_{\text{obs}}(z, \hat{\textbf{n}}) =\, &5\, \text{log}_{10}(1+\delta d_{\text{L}}(z, \hat{\textbf{n}})) \simeq \frac{5}{\ln\,10}\,\delta d_{\text{L}}(z, \hat{\textbf{n}})\,,
\end{align}
where we have assumed that the fluctuation in luminosity distance is small. The variance in the luminosity distance can then be written as 
\begin{align}
    \left \langle \delta d^2_{\text{L}}(z, \hat{\textbf{n}})\right \rangle = \left \langle \delta d^2_{\text{L,lens}}(z, \hat{\textbf{n}})\right \rangle+\left \langle \delta d^2_{\text{L,vel}}(z, \hat{\textbf{n}})\right \rangle\,.
\end{align}
The cross-correlation between lensing and peculiar velocities vanishes because peculiar velocities are integrated along the l.o.s. due to the lensing kernel and so average out to zero~\citep{Hui:2005nm}. The cross-correlation between the l.o.s. peculiar velocities of the host galaxy and the observer is negligible compared to the variance of the peculiar velocity of the host galaxy~\citep{Hui:2005nm}. We choose the redshift bins to be small enough that the variances do not change appreciably within each bin. The lensing contribution to the variance is then given as 
\begin{align}\label{eq:lens_var}
    \nonumber\sigma_{\text{lens}}^2(z, \hat{\textbf{n}}) \equiv & \Big[\frac{5}{\ln 10}\Big]^2\left \langle \delta d^2_{\text{L,lens}}(z, \hat{\textbf{n}})\right \rangle \\
    \nonumber=& \Big[\frac{15 H^2_0 \Omega_{m0}}{2\,\ln 10}\Big]^2\int_{0}^{\chi_s}d\chi \Big[\frac{\chi(\chi_s-\chi)}{\chi_s}\Big]^2 \\
    &\times(1+z)^2\int \frac{dk}{2\pi}k P_{\text{nl}}(k, z)\,,
\end{align}
where we have used Limber's approximation (see Appendix D of~\cite{Hui:2005nm} for more details) and $P_{\text{nl}}(k,z)$ is the nonlinear matter power spectrum at redshift $z$. The velocity contribution is given by
\begin{align}\label{eq:vel_var}
   \nonumber\sigma_{\text{vel}}^2(z, \hat{\textbf{n}}) \equiv & \Big[\frac{5}{\ln 10}\Big]^2\left \langle \delta d^2_{\text{L,vel}}(z, \hat{\textbf{n}})\right \rangle  \\
   \nonumber &= \Big[\frac{5}{\ln 10}\Big]^2\Big[1-\frac{1}{a_s H_s\chi_s}\Big]^2\\
    &\times\int \frac{dk}{6\pi^2} [D^{'}(k, z)]^2 P_{\text{nl}}(k, z = 0)\,,
\end{align}
where $D'(k,z) \equiv -H(z) \frac{d\,D(k,z)}{dz}$ with $D(k,z)$ the linear growth factor for matter fluctuations. Note that it is a function of wavenumber because of nonzero neutrino mass. 

\begin{figure*}[t]
\begin{center}
    \includegraphics[width = 0.50\textwidth]{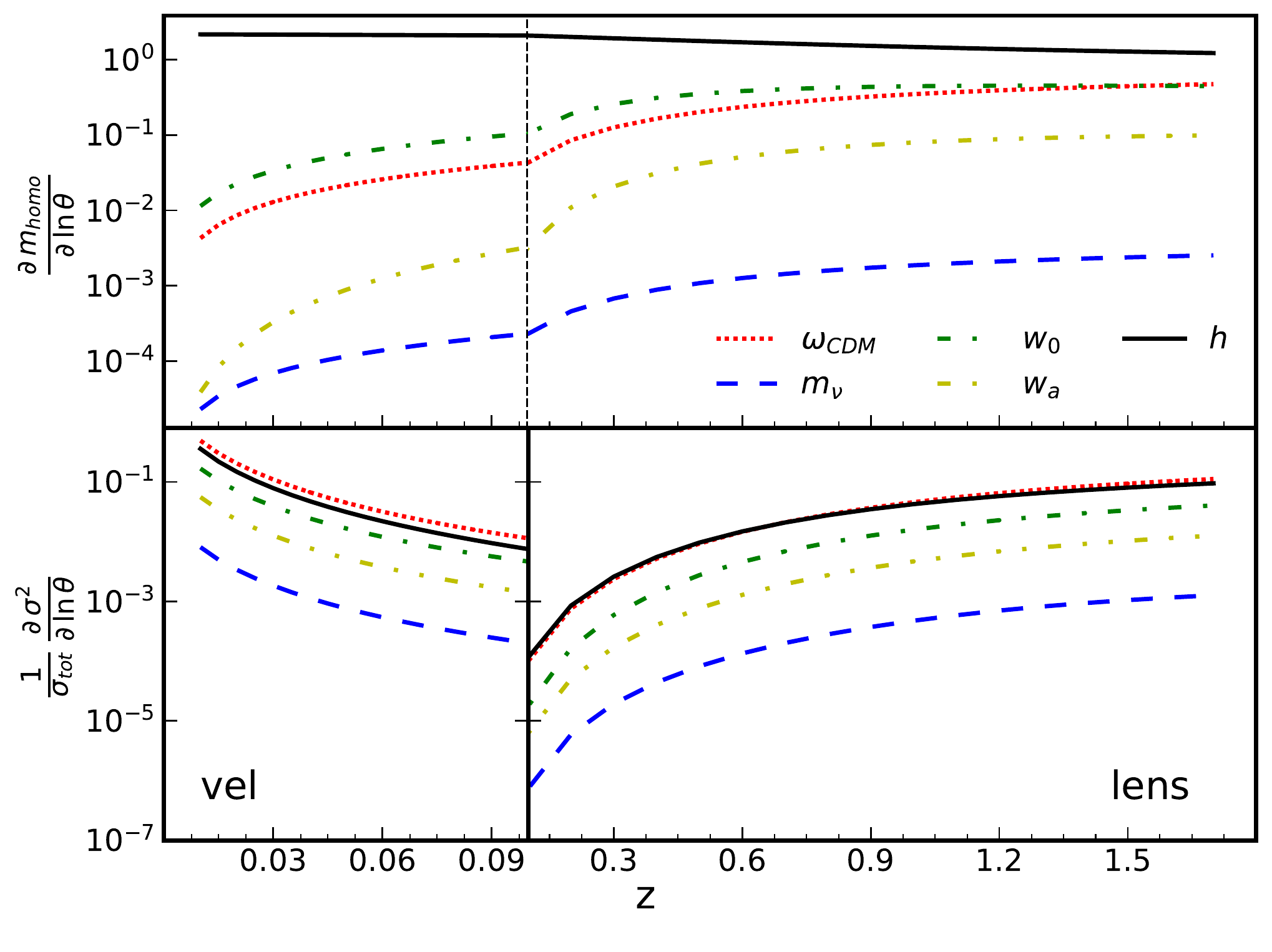}
    \includegraphics[width = 0.48\textwidth]{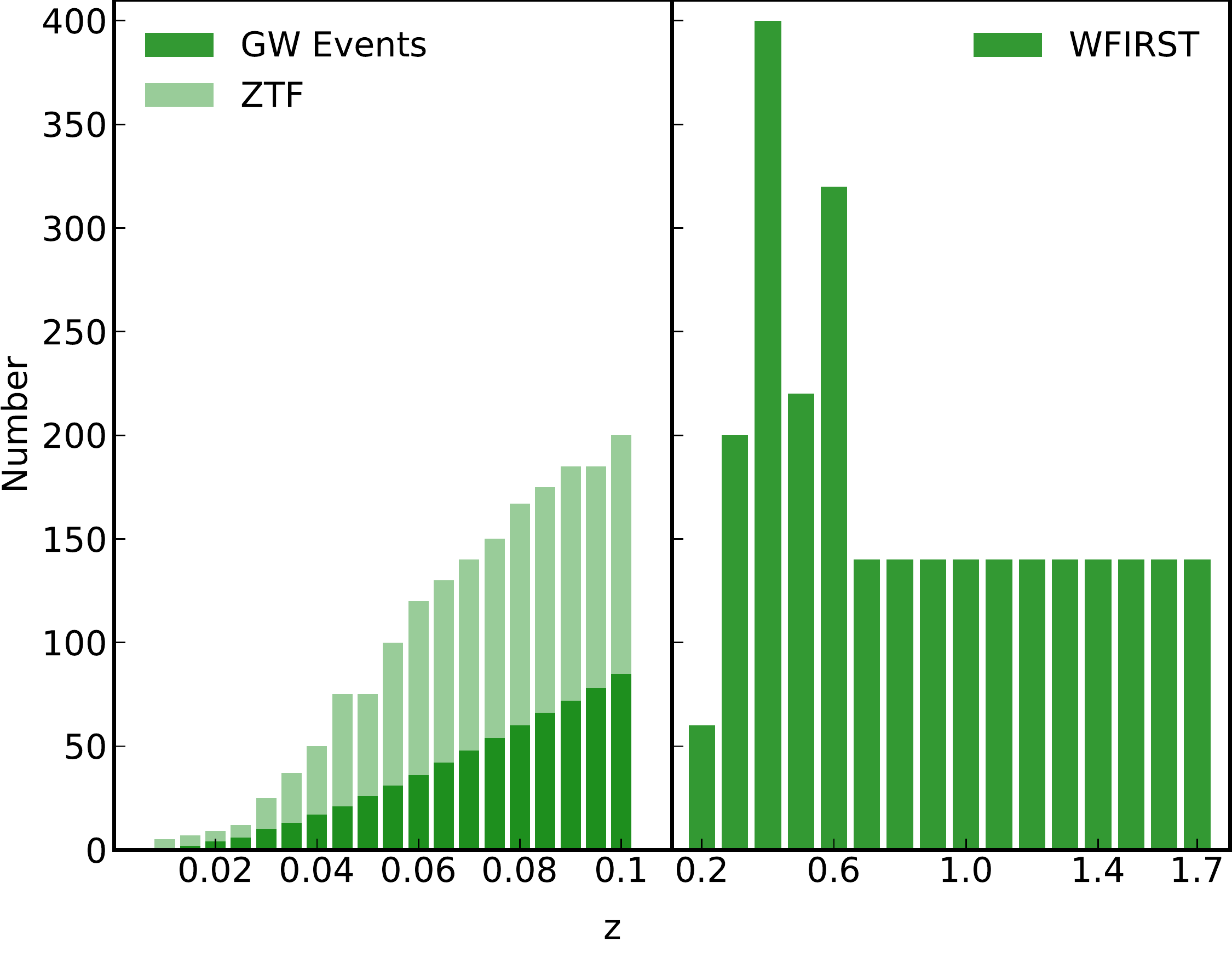}
\end{center}
\vspace*{-0.4cm}
    \caption{(Left) Logarithmic derivatives of the predicted apparent magnitude, $m_{\text{homo}}$ (top), (scaled) velocity variance $\sigma^2_{\text{vel}}$ (bottom left), and (scaled) lensing variance $\sigma^2_{\text{lens}}$ (bottom right), w.r.t. the five cosmological parameters of interest, $\omega_{\text{CDM}}$ (red dotted), sum of neutrino masses $m_{\nu}$ (blue dashed), dark energy equation of state parameters $w_0$ (green dotted--dashed) and $w_a$ (yellow dotted--dotted--dashed), and the Hubble constant $h$ (black solid). Note that for $w_a$ the standard derivative that has been shown as a logarithm of $0$ is not defined. The black dashed line in the upper panel corresponds to $z = 0.1$, which is the highest redshift for our low-$z$ sample, and below which the velocity variance dominates over lensing. (Right) Expected distribution of standard sirens and SNe Ia from {ZTF~\citep{Graham:2019qsw} (left)} and SNe Ia from WFIRST~\citep{Spergel:2015sza} (right).}
    \label{fig:ders}
\end{figure*}

We use the Fisher matrix formalism as described in~\cite{Agrawal:2019yed} to make forecasts for cosmological constraints using standard sirens and candles. The likelihood is taken to be Gaussian, and different events are assumed to be independent of each other. The resultant Fisher matrix becomes~\citep{Agrawal:2019yed}
\begin{align}\label{eq:tot_l}
    \nonumber\mathcal{F}_{ab, \text{tot}} = \sum_{i=1}^{N} &\Bigg[\frac{1}{\sigma^2_{\text{tot}}}\frac{\partial m_{\text{homo}}(z_i)}{\partial \theta_a}\frac{\partial m_{\text{homo}}(z_i)}{\partial \theta_b}\\
    &+\frac{1}{2\,\sigma^4_{\text{tot}}}\frac{\partial \sigma^2_{\text{tot}}(z_i)}{\partial \theta_a}\frac{\partial \sigma^2_{\text{tot}}(z_i)}{\partial \theta_b}\Bigg]\,.
\end{align}
where the sum runs over different observed events, $\theta_a$ denotes the cosmological parameter of interest, $m_{\text{homo}}$ is the magnitude predicted in a homogeneous universe and $\sigma^2_{\text{tot}}$ is the total variance of the observed magnitude, which is given by the quadrature sum of the lensing, peculiar velocity and intrinsic contributions, $\sigma^2_{\text{tot}} = \sigma^2_{\text{lens}}+\sigma^2_{\text{vel}}+\sigma^2_{\text{in}}\,.$ We use $\sigma_{\text{in}} = 0.02$ for GW sources~\citep{Holz:2005df} and $\sigma_{\text{in}} = 0.12$ for SNe Ia~\citep{Spergel:2015sza}.

The top panel of~\reffig{ders} shows the absolute value of derivatives of $m_{\text{homo}}$, $\sigma^2_{\text{lens}}$, and $\sigma^2_{\text{vel}}$ that enter in the Fisher matrix,~\refeq{tot_l}, w.r.t. five cosmological parameters that we wish to constrain---$h$, cold dark matter density at $z=0$, $\omega_{CDM} \equiv \Omega_{CDM} h^2$, $w_0$, $w_a$, and the sum of neutrino masses that we denote by $m_{\nu}$. The variances have been scaled by the total variance (with $\sigma_{\text{in}} = 0.12$) to make them equivalent to the homogeneous magnitude in~\refeq{tot_l}. The first thing to notice is the relative sizes of these derivatives. At low redshifts the derivative of the homogeneous magnitude dominates the lensing contribution for all parameters, whereas at higher redshifts the derivatives of the lensing variance start to become comparable to that of the homogeneous magnitude. In contrast, for all parameters other than $h$ the velocity contribution dominates at the lowest redshifts, but it drops quite sharply as redshift increases. We have verified that the constraints are in fact dominated by the homogeneous part at low redshifts. However, including variances helps to break degeneracies among parameters, and to put much tighter constraints on the dark energy equation of state and sum of neutrino masses as shown in~\cite{Agrawal:2019yed} and \cite{Hada:2018ybu, Hada:2016dje}. 

The derivative of the homogeneous magnitude w.r.t. $h$ is almost an order of magnitude higher than that w.r.t. other parameters, and is $\sim 2$. Thus, at the lowest redshift, even without including information from the lensing or velocity variance, we obtain a fractional error on $h$ (with all other parameters fixed) of $\sim 1$\% from just one GW event of the precision considered. This, while being exceptionally promising, should not be surprising since a perfect measurement of the magnitude corresponds to a perfect measurement of the expansion rate, which at $z \sim 0$ is just the Hubble constant. 

\begin{figure*}[t]
\begin{center}
    \includegraphics[width = 0.49\textwidth]{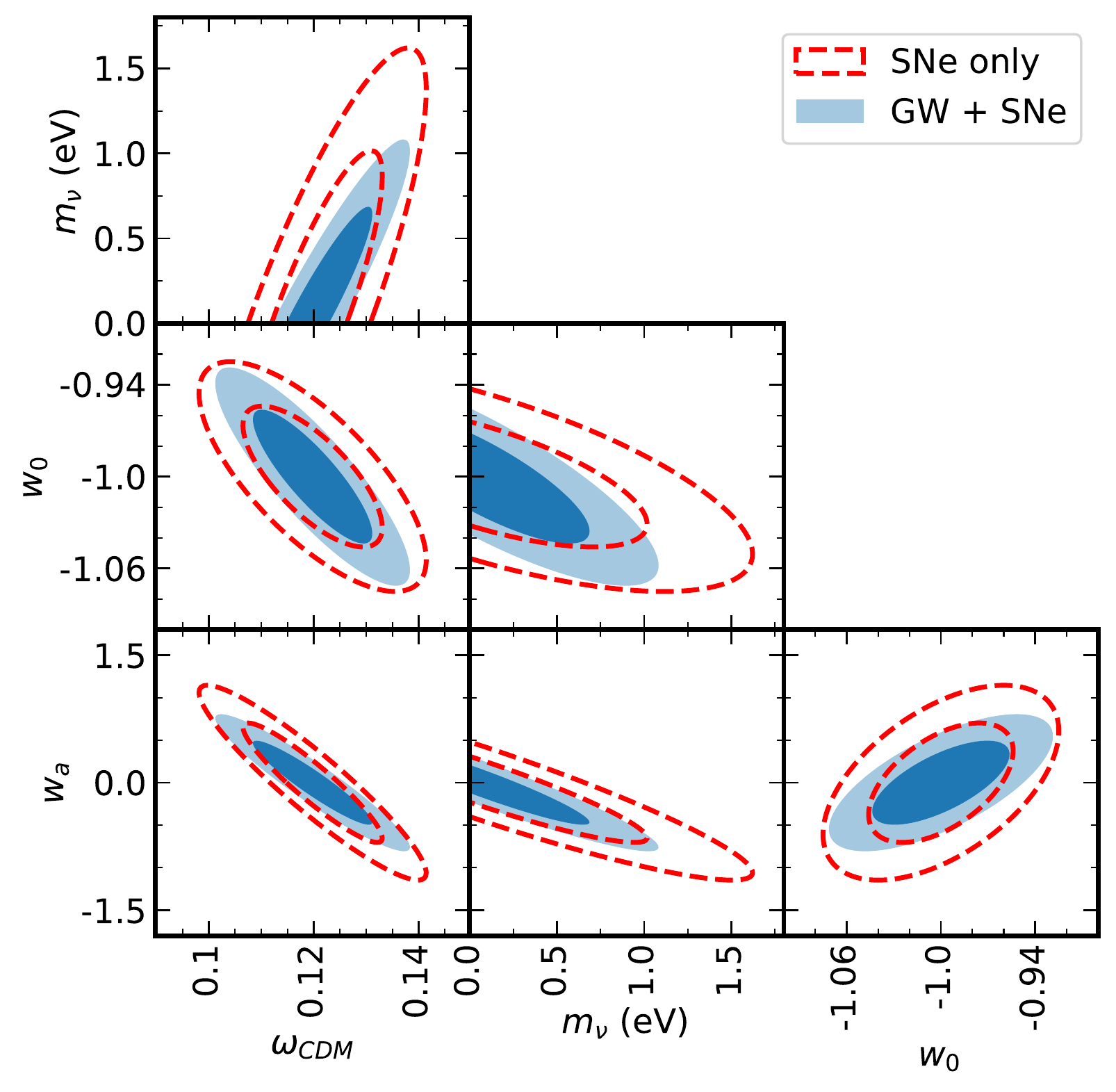}
    \includegraphics[width = 0.49\textwidth]{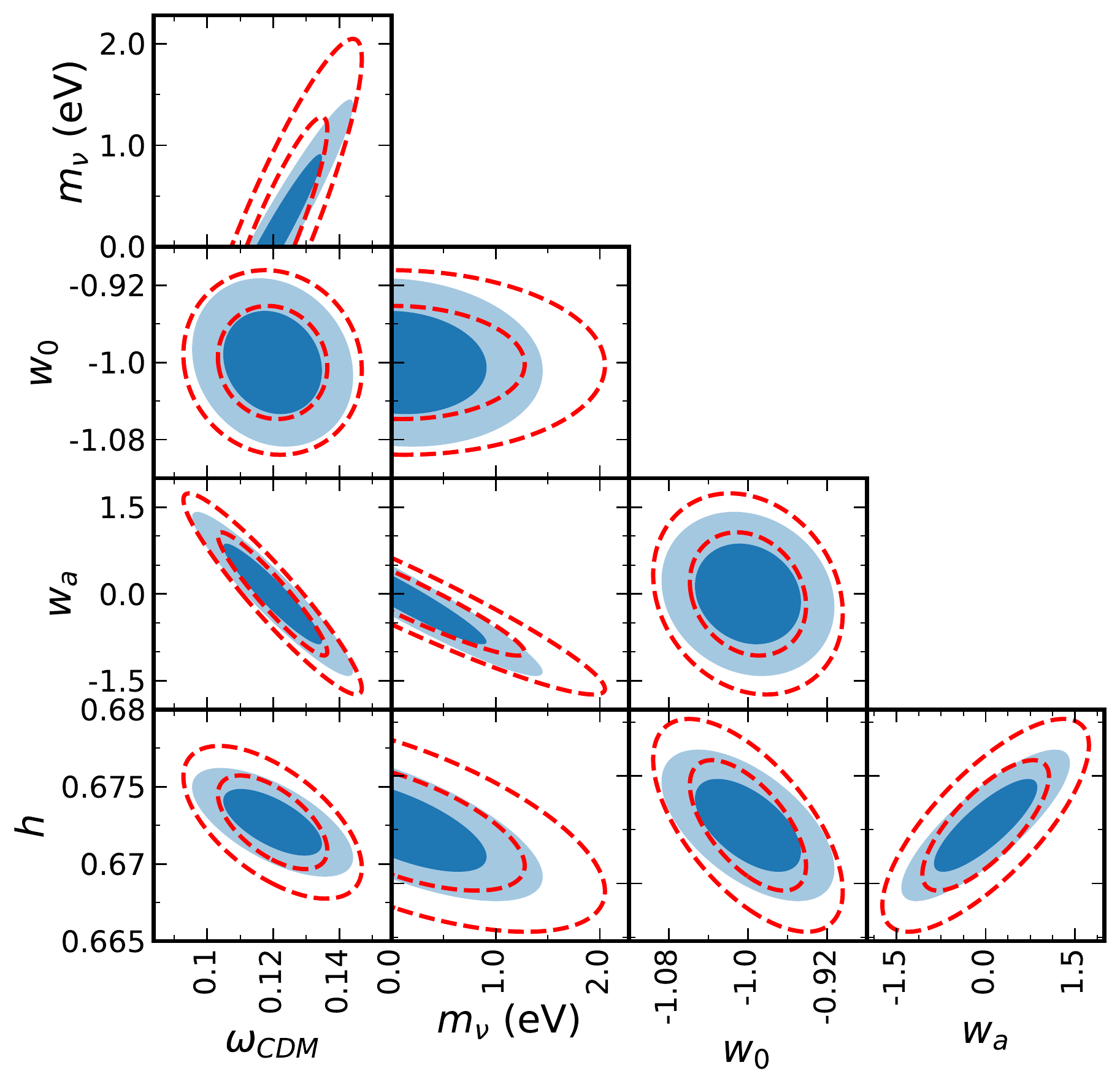}
\end{center}
\vspace*{-0.4cm}    
    \caption{(Left) Joint constraints on $\omega_{\text{CDM}}$, $m_{\nu}$, $w_0$, and $w_a$ with all other cosmological parameters fixed. Red dashed lines denote constraints obtained when using high-$z$ supernovae from WFIRST and low-$z$ supernovae from ZTF, while blue filled regions show constraints obtained when supernovae are combined with low-$z$ GW events. (Right) Same as the left panel, but with $h$ also varying now. {As shown, a constraint of $\sim 0.3$\% on $h$ can be obtained by combining GW sources with SNe Ia, even when the sum of neutrino masses and time-varying dark energy equation of state are simultaneously varied.}}
    \label{fig:4pars}
\end{figure*}

\section{Results}\label{sec:results}

Future experiments such as the Einstein Telescope~\citep{Punturo:2010zz} will observe GW events from sources such as binary black hole (BH-BH), black hole-neutron star (BH-NS) or neutron star-neutron star (NS-NS) mergers. Of these, electromagnetic radiation is emitted along with gravitational radiation for BH-NS and NS-NS mergers. By optical follow-up a redshift can be measured for these events~\citep{TheLIGOScientific:2017qsa}. Alternatively, one can cross-correlate measurements of GW events in luminosity distance space with measurements of galaxy distribution in the same space, to obtain the most likely standard siren redshift~\citep{Kopparapu:2007ib}. Using these redshifts in combination with the precise measurement of luminosity distance for these standard sirens provides extremely tight constraints on fundamental cosmology as we now demonstrate. 

Future experiments will observe $2810$ GW events per $\text{Gpc}^3$~\citep{Abbott:2020uma}. We assume that they are uniformly distributed with this number density out to $z = 0.1$ and that redshifts for these events can be perfectly measured. We scale this number by $1/(1+z)^2$ to account for the decrease in optical flux with redshift. Then, the number of such events observed in each redshift bin of $\Delta z = 0.005$ out to $z = 0.1$ is shown in the left plot of the right panel of \reffig{ders}. 

For the supernova sample, we use ZTF, which will observe low-redshift supernovae in the range $0.01 \lesssim z \lesssim 0.1$~\citep{Graham:2019qsw} and WFIRST, which will observe high-redshift supernovae in the range $0.2 \lesssim z \lesssim 1.7$~\citep{Spergel:2015sza}. The left plot of the right panel of  \reffig{ders} shows the expected distribution of supernovae in different redshift bins for ZTF as the light green bars. The right plot of the right panel of \reffig{ders} shows the expected distribution of supernovae in different redshift bins for WFIRST. Note that while it is dominated by supernovae in the redshift range $z \leq 0.6$, where lensing effects are subdominant compared to the intrinsic uncertainty, the overall number of supernovae for $z \leq 0.6$ and $z > 0.6$ is approximately the same so that lensing effects must be taken into account.

Using the above distributions in~\refeq{tot_l} we can determine the expected constraints on cosmological parameters using just the supernovae or combining the information from both standard candles and low-redshift standard sirens. Note that for high-redshift events the scatter in luminosity distance due to lensing dominates over that from peculiar motion, while for low-redshift events the one from peculiar motion dominates~\citep{Agrawal:2019yed}. Therefore, we do not include the scatter from lensing at low redshifts and the scatter from peculiar velocities at high redshifts.

The top panel of~\reffig{4pars} shows the constraints obtained using these two probes in the 4-parameter space considered in~\cite{Agrawal:2019yed}. The red dashed contours represent constraints obtained when considering only information from supernovae. We include effects only from lensing due to matter along the l.o.s. for the high-redshift probes as peculiar motion has negligible contribution to the luminosity distance for these redshifts~\citep{Hui:2005nm}. The blue filled regions show constraints obtained when we combine supernovae with low-redshift GW events. Note that for low-redshift probes we include the contribution from peculiar motion alone as lensing is not significant at low redshifts~\citep{Hui:2005nm}. 

\refEq{tot_l} shows that cosmological information in the apparent magnitude neatly splits into contributions coming from the homogeneous universe and inhomogeneous universe, the first and second terms, respectively. 
For the homogeneous contribution, $\omega_{\text{CDM}}$ and $m_{\nu}$ are completely degenerate, both contributing via the total matter density only. Using information about the inhomogeneities allows us to break this degeneracy. Since the homogeneous contribution dominates, the shape and orientation of contours for these parameters are similar but not exactly the same. We have verified that if we only consider the homogeneous contribution, while fixing one of these parameters, then the contours do indeed look the same. 

We find that since the luminosity distance is measured to much higher accuracy for a GW source than for an SN Ia, only about 600 GW events add significant information about the sum of neutrino masses over the 4000 SNe Ia from ZTF and WFIRST. This increased constraining power comes from the larger contribution of peculiar velocities once the intrinsic uncertainty becomes lower. These tight constraints also point to the possibility of allowing other parameters to be free. In particular, with an eye to the Hubble tension we consider letting the Hubble constant, $h$, be free. 

The resulting constraints in the 5-parameter space are shown in the bottom panel of~\reffig{4pars}. We see a drastic reduction in the constraints on neutrino mass, with the $2\sigma$ constraint from including GW events almost as good as the $1\sigma$ constraint from just supernovae. The other significant reduction is seen in $h$ as expected. We find that one can constrain $h$ to better than 0.3\% by combining GW events with the supernova measurements even when the time-varying dark energy equation of state and the sum of neutrino masses are simultaneously varied. A key reason for this reduction when using GWs is that they allow us to measure $h$ with extremely high precision leading to an almost fixed $h$. Thus the blue contours are not much worse than the same in the top panel of~\reffig{4pars}. 

\begin{figure}[t]
    \centering
    \includegraphics[width = 0.48\textwidth]{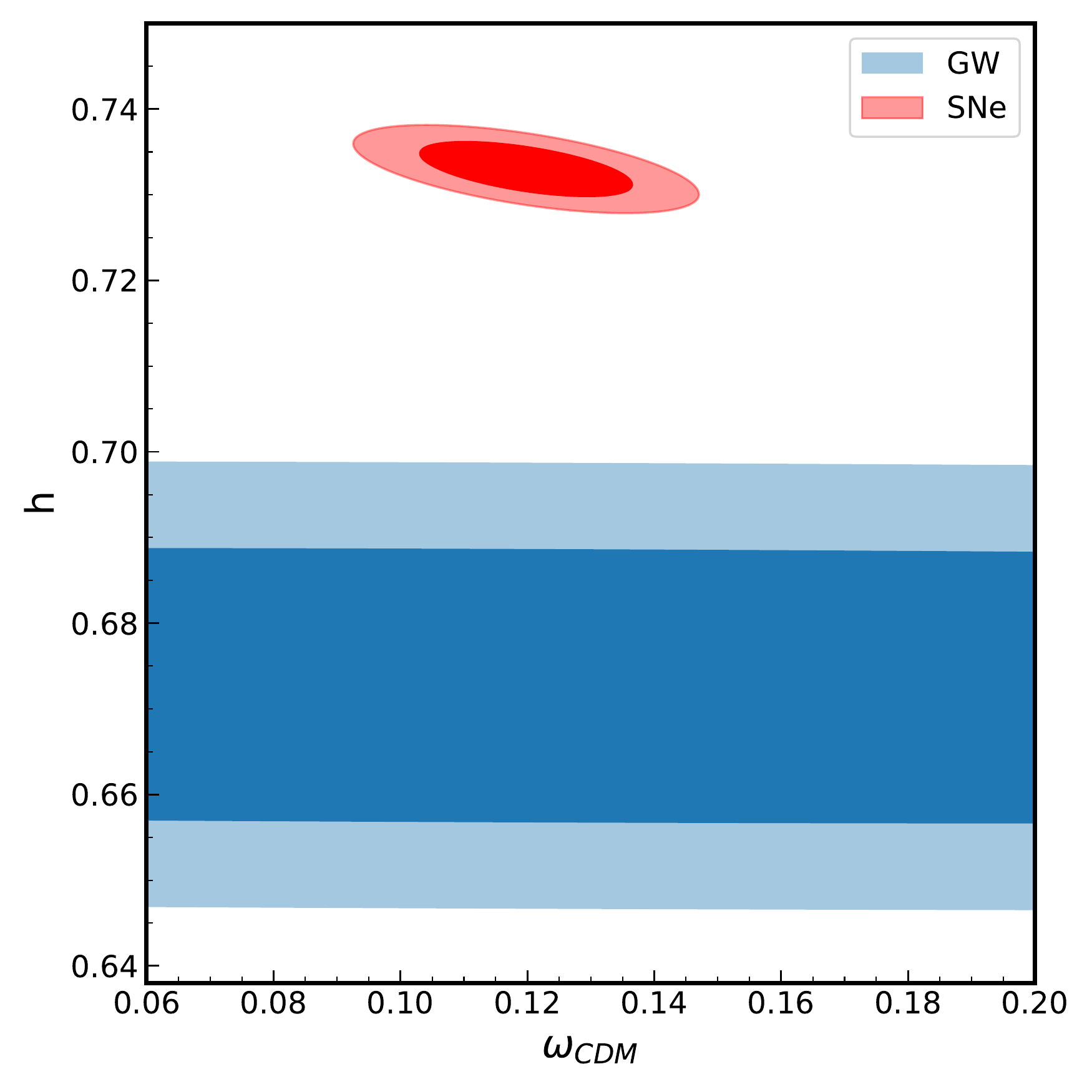}
    \caption{Constraints in $h$-$\omega_{\text{CDM}}$ space after marginalizing over the sum of neutrino masses and dark energy equation of state parameters ($w_0$, $w_a$). Red contours show constraints obtained when using only supernovae from ZTF and WFIRST, while blue contours show constraints obtained when using only low-$z$ GW events. The clear separation of blue and red contours presents an opportunity to resolve the Hubble tension.}
    \label{fig:h_twoz}
\end{figure}

In order to emphasize the promise of GW observations further, we show the constraints on $h$ obtained when marginalizing over all other parameters except $\omega_{\text{CDM}}$ in~\reffig{h_twoz}. The blue contours show constraints obtained with a fiducial value of $h = 0.6727$, consistent with CMB measurements~\citep{ade:2015} and using low-redshift GW events as the probe. The red contours show constraints from SNe Ia from both low (ZTF) and high (WFIRST) redshifts with $h = 0.733$ as the fiducial value, consistent with local measurements~\citep{Verde:2019ivm}. Note that these contours are impossible to draw considering only the homogeneous contribution as $\omega_{\text{CDM}}$ is completely degenerate with $m_{\nu}$ and so no constraints can be obtained if either is marginalized over. It is only with peculiar velocity information that we can constrain one after marginalizing over the other. 

If the Hubble tension arises from unknown systematics, GW and SNe Ia observations should see different values of $h$. In that case it is crucial that we can conclusively say that these two probes are inconsistent with each other, for which the errors on each measurement should be small enough to completely exclude the other. As~\reffig{h_twoz} demonstrates for the probes considered here, this is the case. The two fiducial values we adopt for $h$ are more than $5.5\sigma$ separated from each other (after marginalizing over $\omega_{\text{CDM}}$). Alternatively, if the two contours shown in~\reffig{h_twoz} overlapped, despite different central values, we could not conclude that they are inconsistent with each other and the Hubble tension could be said to still originate from either new physics or unknown systematics. By having nonoverlapping contours we are able to ensure that the difference arises from systematic errors and thus rule out new physics. If, on the other hand, we used $h = 0.71$ and $0.68$ for low and high redshifts, respectively, we still obtain a $2.5\sigma$ separation between the two contours, which can be further improved by observation of a larger number of GW events or by using priors from other measurements. It is also important to note that these conclusions follow even after marginalizing over the sum of neutrino masses or dark energy equation of state parameters, which are poorly constrained by the CMB, or by baryon acoustic oscillation measurements in the absence of a CMB prior. They are thus robust to changing dark energy models whose equation of state can be written in the form $w = w_0 + w_a (1-a)$ or the sum of neutrino masses. 

\section{Conclusion}\label{sec:conclusion}

We have shown that a combination of future GW and SNe Ia observations can resolve the Hubble tension if the universe is assumed to be homogeneous and isotropic on scales spanning the supernovae and GW events we have considered. The numbers considered here can be realized in the next decade. We find that only about 600 GW sources with a measurement of their associated redshifts out to $z\sim 0.1$ are needed for $2\sigma$ constraints on {$h \sim 0.3$\%}. These constraints are obtained even without assuming any prior on the sum of neutrino masses and the dark energy equation of state. Peculiar motion of GW sources is indispensable to break the degeneracy between the sum of neutrino masses and the cold dark matter density in the universe and to obtain tight constraints on the, as yet undetermined, sum of neutrino masses. The Hubble tension presents a fantastic opportunity to test our current understanding of the universe. Future observations made with GWs from inspiraling binary neutron star systems and light from SNe Ia will clarify the origin of the Hubble tension, whether it is coming from new physics or unknown systematics in a homogeneous universe, or that the universe is inhomogeneous.

\acknowledgements
T.O. acknowledges support from the Ministry of Science and Technology of Taiwan under grants No. MOST 106-2119-M-001-031-MY3 and No. MOST 109-2112-M-001-027- and the Career Development Award, Academia Sinica (AS-CDA-108-M02) for the period of 2019 to 2023. T.F. is supported by Grant-in-Aids for Scientific Research from JSPS (No. 17K05453 and No. 18H04357).


\bibliographystyle{apj}

\end{document}